\documentclass[aps,prd,superscriptaddress,nofootinbib]{revtex4}
\usepackage[a4paper, total={6.0in, 9in}]{geometry}
\usepackage{amsfonts,amssymb,latexsym,amsmath} 
\usepackage{bm}
\usepackage{hhline}
\usepackage{longtable}
\usepackage{mathtools} 
\usepackage{array} 
\usepackage{color}
\usepackage{ifthen}
\usepackage{epsfig}
\usepackage{multirow}
\usepackage{enumerate}
\usepackage{dsfont}
\usepackage{psfrag}
\usepackage{float}
\usepackage{appendix}
\usepackage{slashed}
\usepackage{xcolor,booktabs}
\graphicspath{{Figures/}} 
\usepackage{hyperref}
\hypersetup{colorlinks,citecolor=blue,anchorcolor=red,menucolor=red, linkcolor=red,filecolor=red,runcolor=red,urlcolor=blue}

\begin{document}
	
\title{Electromagnetic properties of the $T^+_{cc}$ molecular states}
\author{Ya-Ding Lei}
\email{yadinglei@stumail.nwu.edu.cn}
\affiliation{School of Physics, Northwest University, Xian 710127, China}
\author{Hao-Song Li}
\email{haosongli@nwu.edu.cn}
\affiliation{School of Physics, Northwest University, Xian 710127, China}
\affiliation{Institute of Modern Physics, Northwest University, Xian 710127, China}
\affiliation{Shaanxi Key Laboratory for Theoretical Physics Frontiers, Xian 710127, China}
\affiliation{Peng Huanwu Center for Fundamental Theory, Xian 710127, China}

\begin{abstract}
	In this work, we discuss the electromagnetic properties of the $S$-wave and $D$-wave $T^+_{cc}$ molecular states, which include the magnetic moments, transition magnetic moments and radiative decay widths. According to our results, the magnetic moment of $T^+_{cc}$ state observed experimentally is $-0.09\mu_N$. Meanwhile, we also discuss the relations between the transition magnetic moments of the $S$-wave $T^+_{cc}$ molecular states and the radiative decay widths, and we analyze the proportionality between the magnetic moments of the $T^+_{cc}$ molecular states. These results provide further information on the inner structure of $T^+_{cc}$ molecular states and deepen the understanding of electromagnetic properties of doubly charmed tetraquarks.
\end{abstract}
\maketitle

\section{Introduction}\label{sec1}

In 2003, the Belle Collaboration observed $X(3872)$, which proved the existence of exotic hadronic states \cite{Belle:2003nnu}. The mass of $X(3872)$ is very close to the mass threshold of $D\bar{D^*}$, which promoted the theoretical studies of $D^{(*)} D^{(*)}$ states. Quantum Chromodynamics (QCD) allows for the existence of multi-quark states, such as tetraquark states and pentaquark states. In 2007, the Bell Collaboration reported the first non-zero-charge $Z_c(4430)^+$ with quark configuration of $cu\bar{c}\bar{d}$ \cite{Belle:2007hrb}. Subsequently, a series of tetraquark states were observed, such as $Z_c(3900)^+$ \cite{BESIII:2013ris,Belle:2013yex}, $Z_c(4020)^+$ \cite{BESIII:2013mhi}. These works aroused the interest of theorists in doubly heavy tetraquarks, and as experimental precision improved, more and more exotic tetraquark states were observed and studied \cite{Ali:2018ifm,Xing:2018bqt,Ali:2018xfq,Francis:2018jyb,Maiani:2019cwl,Fontoura:2019opw,Leskovec:2019ioa,Junnarkar:2018twb,Agaev:2018khe,Yang:2019itm,Tang:2019nwv,Hernandez:2019eox}.

In 2021, the LHCb Collaboration observed the doubly charmed tetraquark state $T^+_{cc}$ composed of four quarks $cc\bar{u}\bar{d}$ and quantum number $J^P=1^+$ \cite{LHCb:2021vvq,LHCb:2021auc}. This exotic state shows a narrow peak in the mass spectrum of the $D^0D^0\pi^+$ meson. The mass of the $T^+_{cc}$ state with respect to the $D^{*+}D^0$ mass threshold and the width are
\begin{eqnarray}
	\delta_{m_{\rm{BW}}}=-273\pm61\pm5^{+11}_{-14}~\rm keV,\nonumber\\
	\Gamma_{m_{\rm{BW}}}=410\pm165\pm43^{+18}_{-38}~\rm keV.\nonumber
\end{eqnarray}
For the doubly charmed tetraquarks, their properties have been studied extensively in the past decades, and there exists abundant experimental and theoretical works \cite{Feijoo:2021ppq,Chen:2021vhg,Qiu:2023uno,Deng:2021gnb,Meng:2021jnw,Wang:2023iaz,Noh:2021lqs,Agaev:2021vur,Achasov:2022onn,Ozdem:2021hmk,Albaladejo:2021vln,Agaev:2022ast,Braaten:2022elw,Abreu:2022lfy,Hu:2021gdg,Huang:2021urd,Azizi:2021aib,Xin:2021wcr,Azizi:2023gzv,Wang:2017uld}.

Whether $T^+_{cc}$ is compact tetraquark states or a molecular states remains debated. For example, in ref. \cite{Yun:2022evm}, the author argue that $T_{cc}^+$ state is more likely to be a molecular state. In ref. \cite{Noh:2023zoq}, the author argue that $T_{cc}^+$ state is highly likely to be in a compact configuration. There are many theoretical studies of the interpretation of compact tetraquark states \cite{Esposito:2021ptx,Esposito:2021vhu}. In addition, the interpretation of molecular state is reasonable \cite{Jia:2022qwr,Li:2012ss,Yan:2021wdl,Ding:2023evu,Ortega:2022efc}, since the mass of the $T^+_{cc}$ state is very close to the $D^{*+}D^0$ mass threshold. In Ref. \cite{Ren:2021dsi}, the authors studied the $DD^*$ hadronic molecule interpretation of $T^+_{cc}$ and calculated the mass and decay width of $T^+_{cc}$ using the one-meson exchange potential model. In Ref. \cite{Ling:2021bir}, the authors assumed that $T^+_{cc}$ is an isoscalar $DD^*$ molecule, and used the effective Lagrangian method to study the partial decay width of $T^+_{cc}\to DD\pi$, which resulted in smaller than the central experimental value of the Breit-Wigner fit. In Ref. \cite{Fleming:2021wmk}, the authors studied the decay of $T^+_{cc}$ in the molecular interpretation using effective field theory, and calculated differential distributions in the invariant mass.

For exotic hadronic state, the magnetic moments and the transition magnetic moments are crucial observable physical quantities, which can be used to distinguish between different configurations, and the electromagnetic properties of exotic hadronic states can also help us obtain information about their geometry.

There are a number of theoretical works around the nature of tetraquark states using different approaches, including lattice QCD \cite{Bicudo:2017szl,Yang:2021hrb,Sadl:2021bme,Padmanath:2022cvl}, constituent quark model \cite{Eichten:2017ffp,Tan:2019knr,Deng:2020iqw,Zhao:2020jvl}, and QCD sum rules \cite{Du:2012wp,Chen:2013aba,Su:2022eun}. The constituent quark model has been widely used to describe the magnetic moments of exotic hadronic state quantitatively \cite{Zhou:2022gra,Gao:2021hmv,Wang:2022tib,Wang:2022nqs}. In Ref. \cite{Wu:2022gie}, the authors studied the ground state of the doubly heavy tetraquarks in the constituent quark model and estimated the magnetic moments of the predicted tetraquark states. In Ref. \cite{An:2022qpt}, the authors studied all possible configurations of the ground fully heavy tetraquark states in the constituent quark model, and discussed the spectroscopy behaviors for the fully heavy tetraquark system. 

In this work, we adopt the constituent quark model to study the electromagnetic properties of the $T^+_{cc}$ molecular states, including the magnetic moments, the transition magnetic moments, and the radiative decay widths, and we discuss the effect of orbital excitation on the magnetic moments.

The structure of this paper is as follows. In Sec. \ref{sec2}, we construct the flavor-spin wave functions of the $T^+_{cc}$ molecular states. In Sec. \ref{sec3}, we calculate the magnetic moments of the $S$-wave channel and $D$-wave channel $T^+_{cc}$ molecular states. In Sec. \ref{sec4}, we calculate the transition magnetic moments and the radiative decay widths of the $S$-wave $T^+_{cc}$ molecular states. Finally, we summarize our work and compare our results with those of other studies in Sec. \ref{sec5}.

\section{Wave function}\label{sec2}

The wave function of the hadronic state $\psi$ can be expressed as
\begin{eqnarray}
	\psi=\phi_{\rm color}\otimes\eta_{\rm flavor}\otimes\xi_{\rm spin}\otimes R_{\rm space},
\end{eqnarray}
where $\phi_{\rm color}$, $\eta_{\rm flavor}$, $\xi_{\rm spin}$, and $R_{\rm space}$ represent the color wave function, the flavor wave function, the spin wave function, and the spatial wave function, respectively.

There are four quarks, $cc\bar{q}_1\bar{q}_2$, in the $T^+_{cc}$ states. With the SU(2) flavor symmetry, we construct the wave functions of $T^+_{cc}$ states, we can obtain the flavor wave functions of the $T^+_{cc}$ states by adding the heavy quark $c$ into the flavor wave functions of the two light quarks according to the structures $(Q_1\bar{q}_1)(Q_2\bar{q}_2)$ in the molecular model. Here, $Q$ and $q$ denote the heavy quark and the light quark, respectively. Meanwhile, not all $T^+_{cc}$ molecular states with combinations of spin and isospin exist. Due to Bose-Einstein symmetry, for the $DD$ and $D^*D^*$ states, their quantum numbers $(I+J)$ need to be odd \cite{Peng:2023lfw}.

The $T^+_{cc}$ molecular states are composed of $D^{(*)}D^{(*)}$. For example, it is composed of $D^{(*)0}D^{(*)+}$ with $(I,I_3)=(0,0)$, their flavor wave functions are
\begin{eqnarray}
	\eta_{\rm flavor}&=&\frac{1}{\sqrt{2}}(c\bar{u})(c\bar{d})-\frac{1}{\sqrt{2}}(c\bar{d})(c\bar{u})\nonumber\\
	&=&\frac{1}{\sqrt{2}}\left|D^{(*)0}D^{(*)+}\right\rangle-\frac{1}{\sqrt{2}}\left|D^{(*)+}D^{(*)0}\right\rangle.
\end{eqnarray}

Considering the spin wave functions of the $T^+_{cc}$ molecular states, we can obtain the flavor-spin wave functions. For
example, for the $T^+_{cc}$ molecular composed of $D^*D^*$ with $J^P=1^+$, its flavor-spin wave functions can be written as
\begin{eqnarray}
	\left|\psi\right\rangle&=&\left\{\frac{1}{\sqrt{2}}\left|D^{*0}D^{*+}\right\rangle-\frac{1}{\sqrt{2}}\left|D^{*+}D^{*0}\right\rangle\right\}\nonumber\\
	&&\otimes\left\{\frac{1}{\sqrt{2}}\left|1,1\right\rangle\left|1,0\right\rangle-\frac{1}{\sqrt{2}}\left|1,0\right\rangle\left|1,1\right\rangle\right\}.
\end{eqnarray}

The wave functions of the $T^+_{cc}$ states with different isospins are similar. In Table \ref{Table1}, we collect the flavor wave functions $\eta_{\rm flavor}$ and spin wave function $\xi_{\rm spin}$ of $S$-wave $T^+_{cc}$ molecular states. 

\renewcommand\tabcolsep{0.5cm}
\renewcommand{\arraystretch}{1.50}
\begin{table}[!htbp]
	\caption{The flavor wave functions $\eta_{\rm flavor}$ and spin wave functions $\xi_{\rm spin}$ of $S$-wave $D^{(*)}D^{(*)}$ molecular states, where $I$ and $I_3$ represent the isospin and its third components of the $D^{(*)}D^{(*)}$ system, respectively, and $S$ and $S_3$ represent the spin and its third components of the $D^{(*)}D^{(*)}$ system, respectively.}
	\label{Table1}
	\begin{tabular}{c|c|c}
		\toprule[1.0pt]
		\toprule[1.0pt]
		Systems&$\left|I,I_3\right\rangle$&$\eta_{\rm flavor}$\\
		\hline
		\multirow{4}{*}{$D^{(*)}D^{(*)}$}&$\left|0,0\right\rangle$&$\frac{1}{\sqrt{2}}\left|D^{(*)0}D^{(*)+}\right\rangle-\frac{1}{\sqrt{2}}\left|D^{(*)+}D^{(*)0}\right\rangle$\\
		                                &$\left|1,1\right\rangle$&$\left|D^{(*)+}D^{(*)+}\right\rangle$\\
		                                &$\left|1,0\right\rangle$&$\frac{1}{\sqrt{2}}\left|D^{(*)0}D^{(*)+}\right\rangle+\frac{1}{\sqrt{2}}\left|D^{(*)+}D^{(*)0}\right\rangle$\\
		                                &$\left|1,-1\right\rangle$&$\left|D^{(*)0}D^{(*)0}\right\rangle$\\
		\hline
		Systems&$\left|S,S_3\right\rangle$&$\xi_{\rm spin}$\\
		\hline
		$DD$&$\left|0,0\right\rangle$&$\left|0,0\right\rangle\left|0,0\right\rangle$\\
		\hline
		\multirow{3}{*}{$DD^*$}&$\left|1,1\right\rangle$&$\left|0,0\right\rangle\left|1,1\right\rangle$\\
		                      &$\left|1,0\right\rangle$&$\left|0,0\right\rangle\left|1,0\right\rangle$\\
		                      &$\left|1,-1\right\rangle$&$\left|0,0\right\rangle\left|1,-1\right\rangle$\\
		\hline
		\multirow{9}{*}{$D^*D^*$}&$\left|2,2\right\rangle$&$\left|1,1\right\rangle\left|1,1\right\rangle$\\
		                        &$\left|2,1\right\rangle$&$\frac{1}{\sqrt{2}}\left|1,1\right\rangle\left|1,0\right\rangle+\frac{1}{\sqrt{2}}\left|1,0\right\rangle\left|1,1\right\rangle$\\
		                        &$\left|2,0\right\rangle$&$\frac{1}{\sqrt{6}}\left|1,1\right\rangle\left|1,-1\right\rangle+\frac{\sqrt{2}}{\sqrt{3}}\left|1,0\right\rangle\left|1,0\right\rangle+\frac{1}{\sqrt{6}}\left|1,-1\right\rangle\left|1,1\right\rangle$\\
		                        &$\left|2,-1\right\rangle$&$\frac{1}{\sqrt{2}}\left|1,0\right\rangle\left|1,-1\right\rangle+\frac{1}{\sqrt{2}}\left|1,-1\right\rangle\left|1,0\right\rangle$\\
		                        &$\left|2,-2\right\rangle$&$\left|1,-1\right\rangle\left|1,-1\right\rangle$\\
		                        &$\left|1,1\right\rangle$&$\frac{1}{\sqrt{2}}\left|1,1\right\rangle\left|1,0\right\rangle-\frac{1}{\sqrt{2}}\left|1,0\right\rangle\left|1,1\right\rangle$\\
		                        &$\left|1,0\right\rangle$&$\frac{1}{\sqrt{2}}\left|1,1\right\rangle\left|1,-1\right\rangle-\frac{1}{\sqrt{2}}\left|1,-1\right\rangle\left|1,1\right\rangle$\\
		                        &$\left|1,-1\right\rangle$&$\frac{1}{\sqrt{2}}\left|1,0\right\rangle\left|1,-1\right\rangle-\frac{1}{\sqrt{2}}\left|1,-1\right\rangle\left|1,0\right\rangle$\\
		                        &$\left|0,0\right\rangle$&$\frac{1}{\sqrt{3}}\left|1,1\right\rangle\left|1,-1\right\rangle-\frac{1}{\sqrt{3}}\left|1,0\right\rangle\left|1,0\right\rangle+\frac{1}{\sqrt{3}}\left|1,-1\right\rangle\left|1,1\right\rangle$\\
	    \bottomrule[1.0pt]
	    \bottomrule[1.0pt]
	\end{tabular}
\end{table}

\section{Magnetic moment properties}\label{sec3}

In constituent quark model, the magnetic moment of the hadronic state $\mu$ can be written as the sum of the spin magnetic moments $\mu_{\rm spin}$ and orbital magnetic moments $\mu_{\rm orbital}$ of its corresponding constituents. The total magnetic moment can be expressed as
\begin{eqnarray}
	\mu=\mu_{\rm spin}+\mu_{\rm orbital},
\end{eqnarray}
where the spin magnetic moment $\mu_{\rm spin}$ is the sum of the spins of the each constituent and orbital magnetic moment $\mu_{\rm orbital}$ is related to the orbital angular momentum between the constituents.

In the numerical analysis, the magnetic moment $\mu_{{H}}$ can be calculated by inserting the $z$-component of the magnetic moment operator into the corresponding wave function of the hadron, and the magnetic moment can be expressed as
\begin{eqnarray}
	\mu_{{H}}&=&\left\langle\psi_H\left|\hat{\mu}_z\right|\psi_H\right\rangle, \label{EQA}
\end{eqnarray}
where $\left|\psi_H\right\rangle$ is the wave function corresponding to the hadronic state.

When only considering the $S$-wave channel, there only exists the spin magnetic moment, and the spin operator can be expressed as:
\begin{eqnarray}
\hat{\mu}_{\rm spin}&=&\sum_{i}\frac{e_i}{2m_i}\hat{\sigma}_{i},
\end{eqnarray}
where $e_i$, $m_i$ and $\sigma_{i}$ represent the charge, mass and Pauli’s spin operator of the $i$-th constituent of the hadron $H$, respectively.

In our calculations, we consider the following $S$-wave and $D$-wave channel $T^+_{cc}$ molecular states
\begin{eqnarray}
	D^*D^*[J^P=0^+]&:&\left|^{1}S_0\right\rangle ,\left|^5D_0\right\rangle,\nonumber\\
	D^*D^*[J^P=1^+]&:&\left|^3S_1\right\rangle ,\left|^3D_1\right\rangle ,\left|^5D_1\right\rangle,\nonumber\\
	D^*D^*[J^P=2^+]&:&\left|^5S_2\right\rangle ,\left|^{1}D_2\right\rangle ,\left|^3D_2\right\rangle ,\left|^5D_2\right\rangle.
\end{eqnarray}
Here, we take $\left|^{2S+1}L_J\right\rangle$ to denote the quantum numbers of the corresponding channel, where $S$, $L$, and $J$ represent the quantum numbers of spin angular momentum, orbital angular momentum, and total angular momentum, respectively.

Acrooding to Eq. (\ref{EQA}), the magnetic moments of the $S$-wave $T^+_{cc}$ molecular states can be obtained. Because there are different flavor wave functions with different isospins, the magnetic moments of the $S$-wave $T^+_{cc}$ molecular states with different isospins are different. We take $DD^*$ with $\left|S,S_3\right\rangle=\left|1,1\right\rangle$, $(I,I_3)=(0,0)$ as an example to obtain the magnetic moments  of the $S$-wave $D^*D^*$ molecular state:
\begin{eqnarray}
	\mu_{DD^*}(1^+)&=&\frac{1}{2}(\mu_{D^{*0}}+\mu_{D^{*+}}).
\end{eqnarray}

In the above results, the magnetic moment of the $S$-wave $T^+_{cc}$ molecular state is a linear combination of the magnetic moments of the charmed mesons $D^{(*)}$. Then, we take the charmed meson $D^{*+}$ as an example to obtain the magnetic moment of the charmed meson. In the numerical analysis, we use the constituent quark masses $m_u=0.336~{\rm GeV}$, $m_d=0.336~{\rm GeV}$, $m_c=1.660~{\rm GeV}$ \cite{Wang:2018gpl} to calculate the magnetic moments of $D^{(*)}$ mesons, the flavor-spin wave function can be written as
\begin{eqnarray}
	\chi^{\left|1,1\right\rangle}_{D^{*+}}=\left|c\bar{d} \right\rangle\otimes\left|\uparrow \uparrow  \right\rangle,
\end{eqnarray}
where $\chi^{s}_{f}$ is the flavor-spin wave function of the calculated hadron, the superscript $s$ and the subscript $f$ represent the spin and flavor wave functions, respectively, $\uparrow$ means the third component of the quark spin is $1/2$. According to Eq. (\ref{EQA}), inserting the spin magnetic moment operator $\hat {\mu}_{\rm spin}$ into the corresponding flavor-spin wave function, we can calculate the magnetic moment of the $D^{*+}$ meson as follows
\begin{eqnarray}
	\mu_{D^{*+}}&=&\left\langle\chi^{\left|1,1\right\rangle}_{D^{*+}}\left|\hat{\mu}_z\right|\chi^{\left|1,1\right\rangle}_{D^{*+}}\right\rangle \nonumber\\
	            &=&\left\langle c\bar{d} \uparrow \uparrow \right|\hat{\mu}_z\left|c\bar{d} \uparrow \uparrow \right\rangle  \nonumber\\
	            &=&\mu_c+\mu_{\bar{d}}.
\end{eqnarray}
The magnetic moment of $D^{*+}$ meson is $\mu_c+\mu_{\bar{d}}$, we can obtain the magnetic moment of $D^{*0}$ by the same method, and we list their results in Table \ref{Table2}.

We can obtain the magnetic moment result of the $DD^*$ state by a linear combination of the $D^*$ meson magnetic moments. In the same way, we calculate the magnetic moments of the $T^+_{cc}$ molecular states with different isospins, then the magnetic moments of the $S$-wave $T^+_{cc}$ molecular states are listed in Table \ref{Table3}.

\renewcommand\tabcolsep{0.5cm}
\renewcommand{\arraystretch}{1.50}
\begin{table}[!htbp]
	\caption{The magnetic moment of $S$-wave meson $D^*$, Here, the magnetic moment is in units of the nuclear magnetic moment $\mu_N$.}
	\label{Table2}
	\begin{tabular}{c|c|c}
		\toprule[1.0pt]
		\toprule[1.0pt]
		Mesons&Expressions&Results\\
		\hline
		$D^{*0}$&$\mu_c+\mu_{\bar{u}}$&-1.49\\
		$D^{*+}$&$\mu_c+\mu_{\bar{d}}$&1.31\\
		\bottomrule[1.0pt]
		\bottomrule[1.0pt]
   \end{tabular}
\end{table}

\renewcommand\tabcolsep{0.25cm}
\renewcommand{\arraystretch}{1.50}
\begin{table}[!htbp]
	\caption{The $S$-wave channel magnetic moments of the $T^+_{cc}$ molecular states with different flavor representations. The unit is the nuclear magnetic moment $\mu_N$.}
	\label{Table3}
    \begin{tabular}{c|c|c|c}
		\toprule[1.0pt]
		\toprule[1.0pt]
		\multicolumn{4}{c}{$\frac{1}{\sqrt{2}}\left|D^{(*)0}D^{(*)+}\right\rangle-\frac{1}{\sqrt{2}}\left|D^{(*)+}D^{(*)0}\right\rangle$}\\
		\hline
		Systems&$J^P$&Expressions&Magnetic moments\\
		\hline
		$DD^*$&$1^+$&$\frac{1}{2}(\mu_{D^{*0}}+\mu_{D^{*+}})$&-0.09\\
		\hline
		$D^*D^*$&$1^+$&$\frac{1}{2}(\mu_{D^{*0}}+\mu_{D^{*+}})$&-0.09\\
		\hline
		\multicolumn{4}{c}{$\left|D^{(*)+}D^{(*)+}\right\rangle$}\\
		\hline
		Systems&$J^P$&Expressions&Magnetic moments\\
		\hline
		$DD$&$0^+$&0&0\\
		\hline
		$DD^*$&$1^+$&$\mu_{D^{*+}}$&1.31\\
		\hline
		\multirow{2}{*}{$D^*D^*$}&$0^+$&0&0\\
		                        &$2^+$&$2\mu_{D^{*+}}$&2.62\\
		\hline
		\multicolumn{4}{c}{$\frac{1}{\sqrt{2}}\left|D^{(*)0}D^{(*)+}\right\rangle+\frac{1}{\sqrt{2}}\left|D^{(*)+}D^{(*)0}\right\rangle$}\\
		\hline
		Systems&$J^P$&Expressions&Magnetic moments\\
		\hline
		$DD$&$0^+$&0&0\\
		\hline
		$DD^*$&$1^+$&$\frac{1}{2}(\mu_{D^{*0}}+\mu_{D^{*+}})$&-0.09\\
		\hline
		\multirow{2}{*}{$D^*D^*$}&$0^+$&0&0\\
		                        &$2^+$&$\mu_{D^{*0}}+\mu_{D^{*+}}$&-0.18\\
		\hline
		\multicolumn{4}{c}{$\left|D^{(*)0}D^{(*)0}\right\rangle$}\\
		\hline
		Systems&$J^P$&Expressions&Magnetic moments\\
		\hline
		$DD$&$0^+$&0&0\\
		\hline
		$DD^*$&$1^+$&$\mu_{D^{*0}}$&-1.49\\
		\hline
		\multirow{2}{*}{$D^*D^*$}&$0^+$&0&0\\
		                        &$2^+$&$2\mu_{D^{*0}}$&-2.97\\
		\bottomrule[1.0pt]
		\bottomrule[1.0pt]
    \end{tabular}
\end{table}

On the basis of the magnetic moment results obtained in above work, we summarized the following key points.
\begin{itemize}
	\item The magnetic moments of the $S$-wave channel $T^+_{cc}$  molecular states range from $-2.97\mu_N$ to $2.62\mu_N$, $D^*D^*(2^+) $ with $(I,I_3)=(1,-1)$ and $D^*D^*(2^+) $ with $(I,I_3)=(1,1)$ have minimum and maximum magnetic moment, respectively.
	\item The $S$-wave $T^+_{cc}$ molecular states $DD^*(1^+)$ and $D^*D^*(1^+)$ with $(I,I_3)=(0,0)$ have different quark constituents, but their magnetic moments are the same since they have the same total angular momentum.
	\item The $S$-wave channel $T^+_{cc}$ molecular states with same $I(J^P)$ and different $I_3$ quantum numbers have different magnetic moments, since they have different flavor wave functions as listed in Table \ref{Table1}.
	\item The magnetic moments of the $S$-wave channel $DD$ states with $(I,I_3)=(1,1)$, $(I,I_3)=(1,0)$, $(I,I_3)=(1,-1)$ are all zero. At the same isospin, the magnetic moments of $S$-wave channel $D^*D^*(2^+)$ state and $DD^*(1^+)$ state satisfy the relation $\frac{\mu_{D^*D^*}(2^+)}{\mu_{DD^*}(1^+)}$=2, since their magnetic moments are composed of spin magnetic moments and their spin angular momentums are proportional.
\end{itemize}

Analyzing the numerical results in Table \ref{Table3}, since the mass of the $T^+_{cc}$ ground state with spin-parity quantum numbers $J^P=1^+$ and isospin $I=0$ is very close to the $D^{*+}D^0$ mass threshold, the magnetic moment of the $T^+_{cc}$ state observed experimentally is $-0.09\mu_N$.

In previous work, there exists the study of the magnetic moments of compact $T^+_{cc}$ tetraquark states. In Ref. \cite{Azizi:2021aib}, the authors have already discussed the magnetic moments of the compact $T^+_{cc}$ tetraquark with $J^P=1^+$ in diquark-antidiquark picture using the light-cone QCD sum rules. We compare the magnetic moments of the compact $T^+_{cc}$ tetraquark and the $T^+_{cc}$ molecular states states with $I(J^P)=0(1^+)$. The value of the magnetic moment of the compact $T^+_{cc}$ tetraquark state with $I(J^P)=0(1^+)$ is greater than $0.6\mu_N$ and the value of the magnetic moment of the $T^+_{cc}$ molecular state with $I(J^P)=0(1^+)$ is near 0. Thus the different inner structures of doubly charmed tetraquark states lead to different magnetic moments.

Then, we consider the magnetic moments of $D$-wave channel $T^+_{cc}$ molecular states, the total magnetic moment consists of the spin magnetic moment and the orbital magnetic moment, the operator of orbital magnetic moment can be expressed as
\begin{eqnarray}
	\hat{\mu}_{\rm orbital}=\mu^{L}_{\alpha\beta}\hat{L}_z=(\frac{m_\alpha\mu_\beta}{m_\alpha+m_\beta}+\frac{m_\beta\mu_\alpha}{m_\alpha+m_\beta})\hat{L}_z,
\end{eqnarray}
where the subscripts $\alpha$ and $\beta$ are constituent mesons, and we adopt the constituent mesons masses $m_{D^0}=1864.84~{\rm MeV}$, $m_{D^+}=1869.65~{\rm MeV}$, $m_{D^{*0}}=2006.85~{\rm MeV}$, $m_{D^{*+}}=2010.26~{\rm MeV}$ \cite{ParticleDataGroup:2022pth}.

The spin-orbital wave function $\left|^{2S+1}L_J\right\rangle$ can be expanded using the orbital wave function $Y_{L,m_{L}}$ and the spin wave function $\chi_{S,m_{S}}$ as follows:

\begin{eqnarray}
\left|^{2 S+1} L_J\right\rangle=\sum_{m_{L},m_{S}}C_{Lm_{L},Sm_{S}}^{JM}Y_{L,m_{L}}\chi_{S,m_{S}},
\end{eqnarray}
where $C^{JM}_{Lm_{L},Sm_{S}}$ is the Clebsch-Gordan coefficient, then the $D$-wave channel spin-orbital wave function can be expressed as follows
\begin{eqnarray}
	\left|^3D_1\right\rangle&=&\frac{\sqrt{3}}{\sqrt{5}} Y_{2,2}\chi_{1,-1}-\frac{\sqrt{3}}{\sqrt{10}} Y_{2,1}\chi_{1,0}+\frac{1}{\sqrt{10}} Y_{2,0}\chi_{1,1}, \nonumber\\
	\left|^5D_1\right\rangle&=&\frac{1}{\sqrt{5}} Y_{2,2}\chi_{2,-1}-\frac{\sqrt{3}}{\sqrt{10}} Y_{2,1}\chi_{2,0}+\frac{\sqrt{3}}{\sqrt{10}} Y_{2,0}\chi_{2,1} \nonumber\\
	&&-\frac{1}{\sqrt{5}} Y_{2,-1}\chi_{2,2}, \nonumber\\
	\left|^3D_2\right\rangle&=&\frac{\sqrt{2}}{\sqrt{3}} Y_{2,2}\chi_{1,0}-\frac{1}{\sqrt{3}} Y_{2,1}\chi_{1,1}, \nonumber\\
	\left|^5D_2\right\rangle&=&\frac{\sqrt{2}}{\sqrt{7}} Y_{2,2}\chi_{2,0}-\frac{\sqrt{3}}{\sqrt{7}} Y_{2,1}\chi_{2,1}+\frac{\sqrt{2}}{\sqrt{7}} Y_{2,0}\chi_{2,2}, \nonumber\\
	\left|^5D_0\right\rangle&=&\frac{1}{\sqrt{5}} Y_{2,2}\chi_{2,-2}-\frac{1}{\sqrt{5}} Y_{2,1}\chi_{2,-1}+\frac{1}{\sqrt{5}} Y_{2,0}\chi_{2,0} \nonumber\\
	&&-\frac{1}{\sqrt{5}} Y_{2,-1}\chi_{2,1}+\frac{1}{\sqrt{5}} Y_{2,-2}\chi_{2,2}. \label{EQB}
\end{eqnarray}

Combining Eq. (\ref{EQB}) and the obtained results of the magnetic moments of $S$-wave channel as listed in Table \ref{Table2}, the magnetic moments of of $D$-wave channel can be calculated. We take the magnetic moment $\mu_{\left|^3D_1\right\rangle}$ with $(I,I_3)=(0,0)$ as an example, which can be expressed as
\begin{eqnarray}
	\mu_{\left|^3D_1\right\rangle}&=&\frac{3}{5}(-\frac{1}{2}\mu_{D^{*0}}-\frac{1}{2}\mu_{D^{*+}}+2\mu^L_{D^{*0}D^{*+}}) \nonumber\\
	&&+\frac{3}{10}\mu^L_{D^{*0}D^{*+}}+\frac{1}{10}(\frac{1}{2}\mu_{D^{*0}}+\frac{1}{2}\mu_{D^{*+}}) \nonumber\\
	&=&-\frac{1}{4}\mu_{D^{*0}}-\frac{1}{4}\mu_{D^{*+}}+\frac{3}{2}\mu^L_{D^{*0}D^{*+}},
\end{eqnarray}
then we list the magnetic moments of the $D$-wave channel with different isospins in Table \ref{Table4}.

\renewcommand\tabcolsep{0.65cm}
\renewcommand{\arraystretch}{1.50}
\begin{table*}[!htbp]
	\caption{The $D$-wave magnetic moments of the $T^+_{cc}$ states in the molecular model with different flavor representations. The unit is the nuclear magnetic moment $\mu_N$.}
	\label{Table4}
	\begin{tabular}{c|c|c|c|c}
		\toprule[1.0pt]
		\toprule[1.0pt]
		\multicolumn{5}{c}{$\frac{1}{\sqrt{2}}\left|D^{*0}D^{*+}\right\rangle-\frac{1}{\sqrt{2}}\left|D^{*+}D^{*0}\right\rangle$}\\
		\hline
		States&$J^P$&Quantities&Expressions&Results\\
		\hline
		\multirow{2}{*}{$DD^*$}&$1^+$&$\mu_{\left|^3D_1\right\rangle}$&         $-\frac{1}{4}\mu_{D^{*0}}-\frac{1}{4}\mu_{D^{*+}}+\frac{3}{2}\mu^L_{DD^*}$&-0.02\\
		\cline{2-5}
		&$2^+$&$\mu_{\left|^3D_2\right\rangle}$&$\frac{1}{6}\mu_{D^{*0}}+\frac{1}{6}\mu_{D^{*+}}+\frac{5}{3}\mu^L_{DD^*}$&-0.10\\
		\hline
		\multirow{2}{*}{$D^*D^*$}&\multirow{2}{*}{$1^+$}&$\mu_{\left|^3D_1\right\rangle}$&         $-\frac{1}{4}\mu_{D^{*0}}-\frac{1}{4}\mu_{D^{*+}}+\frac{3}{2}\mu^L_{D^{*0}D^{*+}}$&-0.09\\
		&&$\mu_{\left|^5D_1\right\rangle}$&$\frac{1}{4}\mu_{D^{*0}}+\frac{1}{4}\mu_{D^{*+}}+\frac{1}{2}\mu^L_{D^{*0}D^{*+}}$&-0.09\\
		\hline
		\multicolumn{5}{c}{$\left|D^{*+}D^{*+}\right\rangle$}\\
		\hline
		States&$J^P$&Quantities&Expressions&Results\\
		\hline
		\multirow{2}{*}{$DD^*$}&$1^+$&$\mu_{\left|^3D_1\right\rangle}$&$-\frac{1}{2}\mu_{D^{*+}}+\frac{3}{2}\mu^L_{D^+D^{*+}}$&0.29\\
		\cline{2-5}
		&$2^+$&$\mu_{\left|^3D_2\right\rangle}$&$\frac{1}{3}\mu_{D^{*+}}+\frac{5}{3}\mu^L_{D^+D^{*+}}$&1.49\\
		\hline
		\multirow{4}{*}{$D^*D^*$}&$0^+$&$\mu_{\left|^5D_0\right\rangle}$&0&0\\
		\cline{2-5}
		&\multirow{3}{*}{$2^+$}&$\mu_{\left|^{1}D_2\right\rangle}$&$2\mu^L_{D^{*+}D^{*+}}$&2.62\\
		&&$\mu_{\left|^3D_2\right\rangle}$&$\frac{1}{3}\mu_{D^{*+}}+\frac{5}{3}\mu^L_{D^{*+}D^{*+}}$&2.62\\
		&&$\mu_{\left|^5D_2\right\rangle}$&$\mu_{D^{*+}}+\mu^L_{D^{*+}D^{*+}}$&2.62\\
		\hline
		\multicolumn{5}{c}{$\frac{1}{\sqrt{2}}\left|D^{*0}D^{*+}\right\rangle+\frac{1}{\sqrt{2}}\left|D^{*+}D^{*0}\right\rangle$}\\
		\hline
		States&$J^P$&Quantities&Expressions&Results\\
		\hline
		\multirow{2}{*}{$DD^*$}&$1^+$&$\mu_{\left|^3D_1\right\rangle}$&         $-\frac{1}{4}\mu_{D^{*0}}-\frac{1}{4}\mu_{D^{*+}}+\frac{3}{2}\mu^L_{DD^*}$&-0.02\\
		\cline{2-5}
		&$2^+$&$\mu_{\left|^3D_2\right\rangle}$&$\frac{1}{6}\mu_{D^{*0}}+\frac{1}{6}\mu_{D^{*+}}+\frac{5}{3}\mu^L_{DD^*}$&-0.10\\
		\hline
		\multirow{4}{*}{$D^*D^*$}&$0^+$&$\mu_{\left|^5D_0\right\rangle}$&0&0\\
		\cline{2-5}
		&\multirow{3}{*}{$2^+$}&$\mu_{\left|^{1}D_2\right\rangle}$&$2\mu^L_{D^{*0}D^{*+}}$&-0.18\\
		&&$\mu_{\left|^3D_2\right\rangle}$&$\frac{1}{6}\mu_{D^{*0}}+\frac{1}{6}\mu_{D^{*+}}+\frac{5}{3}\mu^L_{D^{*0}D^{*+}}$&-0.18\\
		&&$\mu_{\left|^5D_2\right\rangle}$&$\frac{1}{2}\mu_{D^{*0}}+\frac{1}{2}\mu_{D^{*+}}+\mu^L_{D^{*0}D^{*+}}$&-0.18\\
		\hline
		\multicolumn{5}{c}{$\left|D^{*0}D^{*0}\right\rangle$}\\
		\hline
		States&$J^P$&Quantities&Expressions&Results\\
		\hline
		\multirow{2}{*}{$DD^*$}&$1^+$&$\mu_{\left|^3D_1\right\rangle}$&$-\frac{1}{2}\mu_{D^{*0}}+\frac{3}{2}\mu^L_{D^0D^{*0}}$&-0.34\\
		\cline{2-5}
		&$2^+$&$\mu_{\left|^3D_2\right\rangle}$&$\frac{1}{3}\mu_{D^{*0}}+\frac{5}{3}\mu^L_{D^0D^{*0}}$&-1.70\\
		\hline
		\multirow{4}{*}{$D^*D^*$}&$0^+$&$\mu_{\left|^5D_0\right\rangle}$&0&0\\
		\cline{2-5}
		&\multirow{3}{*}{$2^+$}&$\mu_{\left|^{1}D_2\right\rangle}$&$2\mu^L_{D^{*0}D^{*0}}$&-2.97\\
		&&$\mu_{\left|^3D_2\right\rangle}$&$\frac{1}{3}\mu_{D^{*0}}+\frac{5}{3}\mu^L_{D^{*0}D^{*0}}$&-2.97\\
		&&$\mu_{\left|^5D_2\right\rangle}$&$\mu_{D^{*0}}+\mu^L_{D^{*0}D^{*0}}$&-2.97\\
		\bottomrule[1.0pt]
		\bottomrule[1.0pt]
	\end{tabular}
\end{table*}

Comparing the magnetic moments of the $S$-wave channel and $D$-wave channel $T^+_{cc}$ states, we can find that the magnetic moments of the $S$-wave channel and $D$-wave channel $T^+_{cc}$ states are not the same. Due to the orbital excitation of the $D$-wave channel, the magnetic moments of $\left|^3D_1\right\rangle$, $\left|^5D_1\right\rangle$, $\left|^3D_2\right\rangle$, and $\left|^5D_2\right\rangle$ state have an variant of 2\% with respect to the $S$-wave magnetic moments. In the $S$-wave channel, the value of the $\left|^{1}S_2\right\rangle$ state is 0, but the value of the $\left|^{1}D_2\right\rangle$ state is $2\mu^L_{D^*D^*}$ in the $D$-wave channel. The magnetic moment of $DD^*\left|^3D_2\right\rangle$ state is approximately five times the magnetic moment of $DD^*\left|^3D_1\right\rangle$ state. Since the values of the orbital magnetic moment and the magnetic moment of $D$ meson satisfy the proportionality relation. For $DD^*$ state with $(I,I_3)=(1,1)$, the value of the orbital magnetic moment $\mu_{D^+D^{*+}}^L$ is close to one-half the value of the magnetic moment $\mu_{D^{*+}}$. In addition, some different states have the same value. For example, for the $T^+_{cc}$ states with $(I,I_3)=(1,1)$ and $J^P=2^+$, there exist $\left|^{1}D_2\right\rangle$, $\left|^{3}D_2\right\rangle$, $\left|^{5}D_2\right\rangle$ states. They have different expressions but the results of these $D$-wave magnetic moments are same, since the value of the orbital magnetic moment $\mu^L_{D^{*+}D^{*+}}$ is approximately equal to the magnetic moments $\mu_{D^{*+}}$. Therefore, the magnetic moments of these states have the same value.

We discuss the $S$-$D$ wave coupling of $T_{cc}$ molecular states. Taking the $DD$ state with $I(J^P)=0(1^+)$ as an example, we study the $S$-$D$ wave coupling. The wave function can be written as
\begin{eqnarray}
	\left|\psi\right\rangle&\sim&\left|DD^*,I=0,J=1\right\rangle\ \otimes Y_{0,0}R_{S1}(r)\nonumber\\
	&&+\left|D^*D^*,I=0,J=1\right\rangle\ \otimes Y_{0,0}R_{S2}(r)\nonumber\\
	&&+D{\rm -wave \ contribution},
\end{eqnarray}
where $R_{Si}(r)$ is the radial wave function of the $S$-wave $DD$ state. When the cutoff parameter is 1.05~{\rm GeV}, the binding energy relative to the $DD^*$ threshold is 1.24~{\rm MeV} and the corresponding root-mean-square radius is 3.11~{\rm fm} which is comparable to the size of the deuteron. The dominant channel is $DD^*\left|^3S_1\right\rangle$, with a probability 96.39\%, the contribution of the $D^*D^*\left|^3S_1\right\rangle$ state is 2.79\%, the probability of the $D$-wave is around 1\%\cite{Li:2012ss}. Since the contribution of the $D$-wave channel is quite small, the dominant channels are $S$-wave channels($\int dr\ r^2({|R_{S1}|}^2+{|R_{S2}|}^2)=99\%$). The contribution of the $D$-wave channel can be neglected. Thus the influence of $S$-$D$ wave coupling is quite minor.

\section{Transition magnetic moment and radiative decay width}\label{sec4}

In this section, we discuss the transition magnetic moments of the $S$-wave $T^+_{cc}$ molecular states, which can provide an important reference for the study of the radiative decay behavior of $T^+_{cc}$ states. For the transition magnetic moments between these discussed $D^*D^*$ states, the calculation is similar to those used to obtain the magnetic moments of $S$-wave $T^+_{cc}$ states, except that the flavor-spin wave functions of the initial and final states are different. The transition magnetic moment of $S$-wave $T^+_{cc}$ molecular states are obtained by the following equation:
\begin{eqnarray}
	\mu_{H\to {H}'}=\left\langle \psi_{H'}\right|\hat{\mu}_ze^{-i\bm{k}\cdot\bm{r}}\left|\psi_H\right\rangle,
\end{eqnarray}
where $\bm{k}$ is the momentum of the emitted photon. When the momentum of the emitted photon is rather small, the factor $\left\langle R_{i'}\right|e^{-i\bm{k}\cdot\bm{r}}\left|R_{i}\right\rangle$ is approximately equal to 1, the spatial wave functions of the initial and final states do not affect the results of the transition magnetic moment. Thus the above equation can be approximated as
\begin{eqnarray}
	\mu_{H\to {H}'}=\left\langle \psi_{H'}\right|\hat{\mu}_z\left|\psi_H\right\rangle. \label{EQC}
\end{eqnarray}

We take $\mu_{D^*D^*\left|2^+\right\rangle\to DD^*\left|1^+\right\rangle}$ with $(I,I_3)=(1,0)$ as an example to illustrate the the procedure of getting the transition magnetic moments between the $S$-wave $D^{(*)}D^{(*)}$ molecular states. According to Table \ref{Table1}, the flavor-spin wave functions of the $D^*D^*\left|2^+\right\rangle$ and $DD^*\left|1^+\right\rangle$ states with $(I,I_3)=(1,0)$ can be constructed as
\begin{eqnarray}
	\chi_{D^*D^*\left|2^+\right\rangle}&=&\left[\frac{1}{\sqrt{2}}\left|D^0D^{*+}\right\rangle+\frac{1}{\sqrt{2}}\left|D^+D^{*0}\right\rangle\right] \nonumber\\
	&&\otimes\left[\frac{1}{\sqrt{2}}\left|1,1\right\rangle\left|1,0\right\rangle+\frac{1}{\sqrt{2}}\left|1,0\right\rangle\left|1,1\right\rangle\right],\nonumber\\
	\chi_{DD^*\left|1^+\right\rangle}&=&\left[\frac{1}{\sqrt{2}}\left|D^0D^{*+}\right\rangle+\frac{1}{\sqrt{2}}\left|D^+D^{*0}\right\rangle\right]\nonumber\\
	&&\otimes\left|0,0\right\rangle\left|1,1\right\rangle.
\end{eqnarray}
According to Eq. (\ref{EQC}), inserting the magnetic moment operator $\hat{\mu}_z$ into the flavor-spin wave functions of the corresponding initial and final states, we can calculate the transition magnetic moment of the $D^*D^*\left|2^+\right\rangle\to DD^*\left|1^+\right\rangle\gamma$ process
\begin{eqnarray}
	&&\mu_{D^*D^*\left|2^+\right\rangle\to DD^*\left|1^+\right\rangle} \nonumber\\
	&=&\left\langle \chi_{DD^*\left|1^+\right\rangle} \right|\hat{\mu}_z\left|\chi_{D^*D^*\left|2^+\right\rangle}\right\rangle \nonumber\\
	&=&\left\langle \chi^{\left|0,0\right\rangle\left|1,1\right\rangle}_{\frac{1}{\sqrt{2}}\left|D^0D^{*+}\right\rangle+\frac{1}{\sqrt{2}}\left|D^+D^{*0}\right\rangle} \right|\hat{\mu}_z\left|\chi^{\frac{1}{\sqrt{2}}\left|1,1\right\rangle\left|1,0\right\rangle+\frac{1}{\sqrt{2}}\left|1,0\right\rangle\left|1,1\right\rangle}_{\frac{1}{\sqrt{2}}\left|D^{*0}D^{*+}\right\rangle+\frac{1}{\sqrt{2}}\left|D^{*+}D^{*0}\right\rangle}\right\rangle \nonumber\\
	&=&\frac{1}{2\sqrt{2}}\mu_{D^{*+}\to D^+}+\frac{1}{2\sqrt{2}}\mu_{D^{*+}\to D^0}.
\end{eqnarray}

In the above expression, transition magnetic moment of the $S$-wave $T^+_{cc}$ molecular states is linear combinations of the transition magnetic moments between its constituent $D^{(*)}$ mesons. We take the $D^{*+}\to D^+\gamma$ process as example to illustrate the transition magnetic moment between $D^{(*)}$ mesons, we construct their flavor-spin wave functions as follows
\begin{eqnarray}
	\chi^{\left|1,0\right\rangle}_{D^{*+}}&=\frac{1}{\sqrt{2}}\left|c\bar{d} \right\rangle\otimes\left|\uparrow \downarrow+\downarrow\uparrow\right\rangle, \nonumber\\
	\chi^{\left|0,0\right\rangle}_{D^+}&=\frac{1}{\sqrt{2}}\left|c\bar{d} \right\rangle\otimes\left|\uparrow \downarrow-\downarrow\uparrow\right\rangle.
\end{eqnarray}
Then, the transition magnetic moment of the $D^{*+}\to D^+\gamma$ process is
\begin{eqnarray}
	\mu_{D^{*+}\to D^+\gamma}&=&\left\langle\chi^{\left|0,0\right\rangle}_{D^+}\left|\hat{\mu}_z\right|\chi^{\left|1,0\right\rangle}_{D^{*+}}\right\rangle \nonumber\\
	&=&\left\langle \frac{c\bar{d}\uparrow\downarrow-c\bar{d}\downarrow\uparrow}{\sqrt{2}} \left|\hat{\mu}_z\right|\frac{c\bar{d}\uparrow\downarrow+c\bar{d}\downarrow\uparrow}{\sqrt{2}} \right\rangle \nonumber\\
	&=&\mu_c-\mu_{\bar{d}}.
\end{eqnarray}

As a result, the magnetic moment of the $D^{*+}\to D^+\gamma$ process is $\mu_c-\mu_{\bar{d}}$. Using the same method, we can obtain the transition magnetic moments of the $D^{*0}\to D^0\gamma$, and we collect their results in Table \ref{Table5}.

\renewcommand\tabcolsep{0.5cm}
\renewcommand{\arraystretch}{1.50}
\begin{table}[!htbp]
	\caption{Transition magnetic moments of $S$-wave charmed meson $D^*$. Here, the transition magnetic moment is in units of the nuclear magnetic moment $\mu_N$.}
	\label{Table5}
	\begin{tabular}{c|c|c}
		\toprule[1.0pt]
		\toprule[1.0pt]
		Processes&Expressions&Results\\
		\hline
		$D^{*0}\to D^0\gamma$&$\mu_c-\mu_{\bar{u}}$&2.24\\
		$D^{*+}\to D^+\gamma$&$\mu_c-\mu_{\bar{d}}$&-0.55\\
		\bottomrule[1.0pt]
		\bottomrule[1.0pt]
	\end{tabular}
\end{table}

Combining the transition magnetic moments of the $D^{*+}\to D^+\gamma$ and $D^{*0}\to D^0\gamma$ processes linearly , we obtain the transition magnetic moments of the $S$-wave $T^+_{cc}$ molecular states, which are collected in Table \ref{Table6}.

\renewcommand\tabcolsep{0.7cm}
\renewcommand{\arraystretch}{1.50}
\begin{table*}[!htbp]
	\caption{The transition magnetic moments between the $S$-wave $T^+_{cc}$ molecular states with different flavor representations. Here, the magnetic moment is in unit of the nuclear magnetic moment $\mu_N$.}
	\label{Table6}
	\begin{tabular}{c|c|c}
		\toprule[1.0pt]
		\toprule[1.0pt]
		\multicolumn{3}{c}{$\frac{1}{\sqrt{2}}\left|D^{(*)0}D^{(*)+}\right\rangle-\frac{1}{\sqrt{2}}\left|D^{(*)+}D^{(*)0}\right\rangle$}\\
		\hline
		Process&Expression&Result\\
		\hline
		$D^*D^*\left|1^+\right\rangle\to DD^*\left|1^+\right\rangle\gamma$&$-\frac{1}{2\sqrt{2}}\mu_{D^{*+}\to D^+}-\frac{1}{2\sqrt{2}}\mu_{D^{*0}\to D^0}$&-0.60\\
		\hline
		\multicolumn{3}{c}{$\left|D^{(*)+}D^{(*)+}\right\rangle$}\\
		\hline
		Processes&Expressions&Results\\
		\hline
		$DD^*\left|1^+\right\rangle\to DD\left|0^+\right\rangle\gamma$&$\mu_{D^{*+}\to D^+}$&-0.55\\
		$D^*D^*\left|0^+\right\rangle\to DD^*\left|1^+\right\rangle\gamma$&$-\frac{1}{\sqrt{6}}\mu_{D^{*+}\to D^+}$&0.23\\
		$D^*D^*\left|2^+\right\rangle\to DD^*\left|1^+\right\rangle\gamma$&$\frac{1}{\sqrt{2}}\mu_{D^{*+}\to D^+}$&-0.39\\
		\hline
		\multicolumn{3}{c}{$\frac{1}{\sqrt{2}}\left|D^{(*)0}D^{(*)+}\right\rangle+\frac{1}{\sqrt{2}}\left|D^{(*)+}D^{(*)0}\right\rangle$}\\
		\hline
		Processes&Expressions&Results\\
		\hline
		$DD^*\left|1^+\right\rangle\to DD\left|0^+\right\rangle\gamma$&$\frac{1}{2}\mu_{D^{*+}\to D^+}+\frac{1}{2}\mu_{D^{*0}\to D^0}$&0.84\\
		$D^*D^*\left|0^+\right\rangle\to DD^*\left|1^+\right\rangle\gamma$&$-\frac{1}{2\sqrt{6}}\mu_{D^{*+}\to D^+}-\frac{1}{2\sqrt{6}}\mu_{D^{*0}\to D^0}$&-0.34\\
		$D^*D^*\left|2^+\right\rangle\to DD^*\left|1^+\right\rangle\gamma$&$\frac{1}{2\sqrt{2}}\mu_{D^{*+}\to D^+}+\frac{1}{2\sqrt{2}}\mu_{D^{*0}\to D^0}$&0.60\\
		\hline
		\multicolumn{3}{c}{$\left|D^{(*)0}D^{(*)0}\right\rangle$}\\
		\hline
		Processes&Expressions&Results\\
		\hline
		$DD^*\left|1^+\right\rangle\to DD\left|0^+\right\rangle\gamma$&$\mu_{D^{*0}\to D^0}$&2.24\\
		$D^*D^*\left|0^+\right\rangle\to DD^*\left|1^+\right\rangle\gamma$&$-\frac{1}{\sqrt{6}}\mu_{D^{*0}\to D^0}$&-0.91\\
		$D^*D^*\left|2^+\right\rangle\to DD^*\left|1^+\right\rangle\gamma$&$\frac{1}{\sqrt{2}}\mu_{D^{*0}\to D^0}$&1.56\\
		\bottomrule[1.0pt]
		\bottomrule[1.0pt]
\end{tabular}
\end{table*}

Analyzing the above results for the transition magnetic moments of $S$-wave $T^+_{cc}$ molecular states, we summarize several key points:
\begin{itemize}
	\item The transition magnetic moments of the $T^+_{cc}$ molecular states can be expressed as linear combinations of the transition magnetic moments of the corresponding constituents, for example the transition magnetic moments of the process $D^*D^*\left|1^+\right\rangle\to DD^*\left|1^+\right\rangle\gamma$ consists of transition magnetic moments of $D^{*+}\to D^+\gamma$ process and $D^{*0}\to D^0\gamma$ process.
	\item The maximum transition magnetic moment is $2.24\mu_N$, corresponding to the process of $DD^*\left|1^+\right\rangle\to DD\left|0^+\right\rangle\gamma$ with $(I,I_3)=(1,-1)$, and the minimum transition magnetic moment is $-0.91\mu_N$, corresponding to the processes  of $D^*D^*\left|0^+\right\rangle\to DD^*\left|1^+\right\rangle\gamma$ with $(I,I_3)=(1,-1)$.
	\item For radiative decay processes with $I=1$, their transition magnetic moments satisfy some proportional relations, for example
	\begin{eqnarray}
		\frac{\mu_{D^*D^*\left|2^+\right\rangle\to DD^*\left|1^+\right\rangle\gamma}}{\mu_{D^*D^*\left|0^+\right\rangle\to DD^*\left|1^+\right\rangle\gamma}}&=&-\sqrt{3},\nonumber\\
		\frac{\mu_{DD^*\left|1^+\right\rangle\to DD\left|0^+\right\rangle\gamma}}{\mu_{D^*D^*\left|2^+\right\rangle\to DD^*\left|1^+\right\rangle\gamma}}&=&\sqrt{2}.\nonumber
	\end{eqnarray}
\end{itemize}

The radiative widths of the $S$-wave $T^+_{cc}$ molecular states can provide important information for exploring its inner structure, so the radiative width is also a crucial physical quantity, and the radiative decay width is closely related to the transition magnetic moment. Based on the transition magnetic moments obtained above, we can calculate the radiative decay width of the S-wave $T^+_{cc}$ molecular state. The relationship between the radiative decay width $\Gamma_{H\to H^{'}\gamma}$ and the transition magnetic moment can be expressed as
\begin{eqnarray}
	\Gamma_{H \to H^{\prime}\gamma}=\begin{cases}
		\alpha_{\rm{EM}}\frac{E_{\gamma}^3}{3m_p^2} \frac{J_{H}+1}{J_{H}}\frac{\left|\mu_{H \to H^{\prime}}\right|^2}{\mu_N^2}     & \text{,}~~J_H=J_{H^{\prime}}, \\
		\alpha_{\rm{EM}}\frac{E_{\gamma}^3}{3m_p^2} J_{H}\frac{\left|\mu_{H \to H^{\prime}}\right|^2}{\mu_N^2}     & \text{,}~~J_H=J_{H^{\prime}}+1,\\
		\alpha_{\rm{EM}}\frac{E_{\gamma}^3}{3m_p^2} \frac{J_{H^{\prime}}(2J_{H^{\prime}}+1)}{2J_{H}+1}\frac{\left|\mu_{H \to H^{\prime}}\right|^2}{\mu_N^2}     & \text{,}~~J_H=J_{H^{\prime}}-1.
	\end{cases}\nonumber\\
\end{eqnarray}

In the above equation, the electromagnetic fine structure constant $\alpha_{\rm{EM}}$ is $1/137$, $m_p$ is the the mass of proton with $m_p=0.938~{\rm GeV}$\cite{Chen:2013aba}. For the $H\to H^{'}\gamma$ process, $E_{\gamma}$ is the momentum of the emitted photon, which can be written as 
\begin{eqnarray}
E_{\gamma}=\frac{m^2_H-m^2_{H^\prime}}{2m_H},
\end{eqnarray}
where $H$ and $H^\prime$ represent the tetraquark molecular states we discuss, and $m_H$ and $m_{H^\prime}$ represent the masses of the corresponding hadrons, respectively. In Table \ref{Table7}, the expressions and results of the $S$-wave $T^+_{cc}$ molecular states are collected.

As presented in Table \ref{Table7}, the effect of the transition magnetic moments on the radiative decay width of the $S$-wave $T^+_{cc}$ molecular states is significant. For example, the $D^*D^*\left|2^+\right\rangle\to DD^*\left|1^+\right\rangle\gamma$ process has a larger radiative decay width than the $D^*D^*\left|0^+\right\rangle\to DD^*\left|1^+\right\rangle\gamma$ process, since the $D^*D^*\left|2^+\right\rangle\to DD^*\left|1^+\right\rangle\gamma$ process has a larger transition magnetic moment than the $D^*D^*\left|0^+\right\rangle\to DD^*\left|1^+\right\rangle\gamma$ process. The radiative decay width of the $DD^*\left|1^+\right\rangle\to DD\left|0^+\right\rangle\gamma$ with $(I,I_3)=(1,-1)$ is much larger than that of $DD^*\left|1^+\right\rangle\to DD\left|0^+\right\rangle\gamma$ with $(I,I_3)=(1,1)$, which is similar to the radiative decay behavior of the $D^{*0}\to D^0\gamma$ and $D^{*+}\to D^+\gamma$ processes. In addition, most of the radiative decay widths with $(I,I_3)=(1,0)$ and $(I,I_3)=(0,0)$ are around $5.0~{\rm keV}$, while the decay width of $D^*D^*\left|0^+\right\rangle\to DD^*\left|1^+\right\rangle\gamma$ process is the less than $3.0~{\rm keV}$, and the radiative decay widths of the processes $DD^*\left|1^+\right\rangle\to DD\left|0^+\right\rangle\gamma$, $D^*D^*\left|0^+\right\rangle\to DD^*\left|1^+\right\rangle\gamma$, $D^*D^*\left|2^+\right\rangle\to DD^*\left|1^+\right\rangle\gamma$ with $(I,I_3)=(1,-1)$ are greater than $10.0~{\rm keV}$. Analyzing the numerical results, it can be noted that the values of the radiative decay widths with different isospins vary considerably.

\renewcommand\tabcolsep{0.7cm}
\renewcommand{\arraystretch}{1.50}
\begin{table}[!htbp]
	\caption{The radiative decay widths between the $S$-wave $T^+_{cc}$ molecular states. Here, the radiative decay width is in units of {\rm keV}.}
	\label{Table7}
	\begin{tabular}{c|c}
		\toprule[1.0pt]
		\toprule[1.0pt]
		\multicolumn{2}{c}{$\frac{1}{\sqrt{2}}\left|D^{(*)0}D^{(*)+}\right\rangle-\frac{1}{\sqrt{2}}\left|D^{(*)+}D^{(*)0}\right\rangle$}\\
		\hline
		Process&Radiative decay width\\
		\hline
		$D^*D^*\left|1^+\right\rangle\to DD^*\left|1^+\right\rangle\gamma$&5.24\\
		\hline
		\multicolumn{2}{c}{$\left|D^{(*)+}D^{(*)+}\right\rangle$}\\
		\hline
		Processes&Radiative decay widths\\
		\hline
		$DD^*\left|1^+\right\rangle\to DD\left|0^+\right\rangle\gamma$&2.23\\
		$D^*D^*\left|0^+\right\rangle\to DD^*\left|1^+\right\rangle\gamma$&1.18\\
		$D^*D^*\left|2^+\right\rangle\to DD^*\left|1^+\right\rangle\gamma$&2.18\\
		\hline
		\multicolumn{2}{c}{$\frac{1}{\sqrt{2}}\left|D^{(*)0}D^{(*)+}\right\rangle+\frac{1}{\sqrt{2}}\left|D^{(*)+}D^{(*)0}\right\rangle$}\\
		\hline
		Processes&Radiative decay widths\\
		\hline
		$DD^*\left|1^+\right\rangle\to DD\left|0^+\right\rangle\gamma$&5.23\\
		$D^*D^*\left|0^+\right\rangle\to DD^*\left|1^+\right\rangle\gamma$&2.63\\
		$D^*D^*\left|2^+\right\rangle\to DD^*\left|1^+\right\rangle\gamma$&5.24\\
		\hline
		\multicolumn{2}{c}{$\left|D^{(*)0}D^{(*)0}\right\rangle$}\\
		\hline
		Processes&Radiative decay widths\\
		\hline
		$DD^*\left|1^+\right\rangle\to DD\left|0^+\right\rangle\gamma$&37.52\\
		$D^*D^*\left|0^+\right\rangle\to DD^*\left|1^+\right\rangle\gamma$&18.81\\
		$D^*D^*\left|2^+\right\rangle\to DD^*\left|1^+\right\rangle\gamma$&36.68\\
		\bottomrule[1.0pt]
		\bottomrule[1.0pt]
    \end{tabular}
\end{table}

\section{Summary}\label{sec5}

The observation of the $T^+_{cc}$ state provides a new platform to study the doubly charmed tetraquark state, theorists have done extensive research around the nature of the doubly charmed tetraquark state. The electromagnetic properties of hadronic molecular states can also help us to further understand their inner structure, but there is no more progress in related research.

In this work, we construct the flavor-spin wave functions of $T^+_{cc}$ molecular states using the constituent quark model. We systematically study their electromagnetic properties, including magnetic moments, transition magnetic moments and radiative decay widths. According to our results, the magnetic moment of the $T^+_{cc}$ state observed experimentally is $-0.09\mu_N$. In addition, we discuss the effect of orbital excitations on the magnetic moments, and we find the $D$-wave magnetic moment have an variant of 2\% with respect to the $S$-wave magnetic moments.

Meanwhile, we also discuss the radiative decay widths, the results show that the radiative decay widths are closely related to the transition magnetic moments, and there are several radiative decay processes of $T^+_{cc}$ molecular states with significant widths. We compare our results with other studies on the natural properties of $T^+_{cc}$ in Table \ref{Table8}. We hope that the present study will inspire more research on the electromagnetic properties of $T^+_{cc}$ molecular states, which can enrich our understanding of the inner structure of hadronic molecular states.

\renewcommand\tabcolsep{0.7cm}
\renewcommand{\arraystretch}{1.50}
\begin{table}[!htbp]
	\caption{Comparison of the quantities of the tetraquark $T^+_{cc}$ state in the article with the results of other studies, including nonrelativistic quark model (NQM)\cite{Deng:2021gnb}, MIT bag model\cite{Zhang:2021yul}, diffusion monte carlo (DMC)\cite{Mutuk:2023oyz}, constituent quark model (CQM)\cite{Wu:2022gie}, one-boson-exchange (OBE)\cite{Ling:2021bir}, effective field theory (EFT)\cite{Fleming:2021wmk}, coupled-channel effective field theory (CCEFT)\cite{Meng:2021jnw}. Here, the magnetic moment is in units of $\mu_N$ and the radiative decay width is in units of {\rm keV}.}
	\label{Table8}
	\begin{tabular}{ccc}
		\toprule[1.0pt]
		\toprule[1.0pt]
		Quantities&$\mu_{T^+_{cc}}$&$\Gamma[T^+_{cc}\to D^+D^0\gamma]$\\
		\hline
		NQM\cite{Deng:2021gnb}&0.13&---\\
		MIT bag model\cite{Zhang:2021yul}&0.88&---\\
		DMC\cite{Mutuk:2023oyz}&0.28&---\\
		CQM\cite{Wu:2022gie}&0.732&---\\
		OBE\cite{Ling:2021bir}&---&10.0\\
		EFT\cite{Fleming:2021wmk}&---&2.8\\
		CCEFT\cite{Meng:2021jnw}&---&5.0\\
		This work&-0.09&5.23\\
		\bottomrule[1.0pt]
		\bottomrule[1.0pt]
	\end{tabular}
\end{table}

\section*{Acknowledgement}

This project is supported by the National Natural Science Foundation of China under Grants No. 11905171. This work is also supported by the Natural Science Basic Research Plan in Shaanxi Province of China (Grant No. 2022JQ-025) and Shaanxi Fundamental Science Research Project for Mathematics and Physics (Grant No.22JSQ016).

\end{document}